\documentstyle[12pt]{article}
\def\lapproxeq{\lower .7ex\hbox{$\;\stackrel{\textstyle <}{\sim}\;$}}

\def\lapproxeq{\lower .7ex\hbox{$\;\stackrel{\textstyle <}{\sim}\;$}}

\begin{document}

\titlepage

\begin{flushright} DTP/93/28 \\ September 1993
\end{flushright}

\begin{center}
\vspace*{2cm}
{\large{\bf Properties of the BFKL equation and structure function \\
predictions for HERA }}
\end{center}

\vspace*{.75cm}
\begin{center}
A.J.\ Askew,  J.\ Kwiecinski\footnote{Permanent address: Department of
Theoretical Physics, H.\ Niewodniczanski Institute of Nuclear Physics, Krakow,
Poland.}, A.D.\ Martin, \\
Department of Physics, University of Durham, DH1 3LE, England.  \\

\end{center}

\begin{center}
and
\end{center}

\begin{center}
P.J.\ Sutton \\
Department of Physics, University of Manchester, M13 9PL, England. \\
\end{center}

\vspace*{1cm}

\begin{abstract}
The general properties of the Lipatov or BFKL equation are reviewed.
Modifications to the infrared region are proposed.  Numerical predictions for
the deep-inelastic electron-proton structure functions at small $x$ are
presented and confronted with recent HERA measurements.
\end{abstract}

\newpage
\noindent {\large{\bf 1.  Introduction}}

Deep-inelastic electron-proton scattering experiments at HERA are measuring
the structure function $F_2(x,Q^2)$ in the previously unexplored small $x$
regime, $10^{-2} < x < 10^{-4}$.  As usual $Q^2 \equiv -q^2$ and $x \equiv
Q^2/2p\cdot q$, where $p$ and $q$ are the four momenta of the incoming proton
and the virtual photon probe respectively.  The deep-inelastic experimental
observables reflect the small $x$ behaviour of the gluon, which is by far the
dominant parton in this kinematic region.  In particular the small $x$
behaviour of $F_2$ is driven by the $g \rightarrow q\bar{q}$ transition.  At
such small values of $x$ soft gluon emission and the associated virtual gluon
corrections give rise to powers of $\alpha_s$log$(1/x)$ which clearly have to
be resummed.  Technically, in the physical gauge, this is equivalent to the
summation of gluon ladder diagrams (with the virtual contributions leading to
so-called gluon Reggeisation [1-6] or, alternatively, to
the introduction of the non-Sudakov form factor [7-9]).
These ladder diagrams are  a universal feature of all small $x$ processes
driven by the gluon.  For instance they occur in the perturbative QCD
description of the structure functions $F_2$ and $F_L$, heavy quark-pair and
$J/\psi$ production, prompt photon production, deep-inelastic diffraction and
deep-inelastic events containing a measured jet.

The resummation of the leading powers of $\alpha_s$log$(1/x)$ leads to the
Lipatov (or BFKL) equation [1-10].  Here we wish to
emphasize two characteristic features of the gluon obtained on solving this
equation.  First the gluon exhibits an $x^{-\lambda}$ growth as $x$ decreases
with $\lambda \sim 0.5$.  This power behaviour of the gluon should manifest
itself in all the small $x$ processes listed above since they incorporate the
universal gluon ladder.  We speak of a \lq\lq hard QCD" or \lq\lq Lipatov"
pomeron with intercept $\alpha_L(0) = 1 + \lambda \sim 1.5$, applicable to
$Q^2$ values sufficiently large to be in the perturbative regime.  This is to
be contrasted with the universal \lq\lq soft" pomeron (associated with $Q^2
\sim 0$) whose intercept has been phenomenologically determined to be
$\alpha_P(0) \sim 1.08$ \cite{PVL}.  The second feature concerns the
transverse momenta, $k_T$, of the gluons along the ladder.  It is no longer
true that the transverse momenta are strongly ordered as is the case in the
\lq\lq large" $x$, large $Q^2$ \lq\lq Altarelli-Parisi" regime.  Rather, as we
evolve down in $x$, we have diffusion arising from a \lq\lq random walk" in
$k_T$ which leads to a broadening of the starting $k_T$-distribution both to
larger and to smaller values of $k_T$.

Instead of simply concentrating on the gluon distribution alone, here
we study the impact of these perturbative QCD effects on the small $x$
behaviour of the structure function $F_2(x,Q^2)$.  However the
implementation of this programme requires knowledge of the gluon for
all $k^2_T$ including the infrared region with $k^2_T$ close to zero.
In this intrinsic non-perturbative or confinement region the BFKL
equation, which is based on perturbative QCD, is not expected to be
valid, and has to be modified.  The major aim of this paper is to
review and to critically examine the solutions of the Lipatov equation
and their impact on the structure function predictions using
physically motivated modifications for small $k^2_T$.  In particular
we study the sensitivity of the predictions to a detailed
parametrization of the infrared region which satisfies the gauge
invariance constraints as $k^2_T \rightarrow 0$.  We find that the
magnitude of $F_2$ has a sizeable uncertainty whereas the
$x^{-\lambda}$ shape remains remarkably stable, with $\lambda$ being
only weakly dependent on the choice of values of the infrared
parameters.  A comparison of the measurements of $F_2$ with the
predicted $x^{-\lambda}$ behaviour therefore provides a test of the
Lipatov perturbative QCD effect.  The Lipatov pomeron may, however,
only give part of the increase of $F_2$ and so we must allow for the
effect of the \lq\lq background" contribution when comparing with the
data.

The $x^{-\lambda}$ growth of the gluon will eventually be suppressed by
shadowing effects.  We shall find these effects are relatively unimportant in
the HERA regime, unless very small gluon concentrations exist within the
proton.

The outline of the paper is as follows.  In section 2 we review the present
knowledge of the properties of the solutions to the BFKL or Lipatov equation.
 In
particular we illustrate the effects of diffusion and of the introduction of
cut-offs in $k_T$.  In section 3 we introduce our infrared modifications of
the BFKL equation and give numerical predictions for the structure
functions.  We confront these predictions with recent HERA data.  Section 4
contains our conclusions.

\vspace*{1cm}
\noindent {\large {\bf 2.  The behaviour of the solutions of the BFKL
equation}}

A central problem in small $x$ physics is the stability of the solutions of
the BFKL or Lipatov equation [1-9] to
contributions from the infrared and ultraviolet regions of the transverse
momenta of the emitted gluons.  This is reflected in the dependence of the
solutions to the choice of the transverse momentum cut-offs.  Several general
properties of the solutions of the Lipatov equation are known, which are
scattered widely in the literature \cite{BL,LIP1,GLR1,LEVORS,BARTL,COLLAND}.
In this section we draw these together and attempt to present a reasonably
self-contained and coherent discussion.
\vspace*{.5cm}

\noindent {\bf (a) Solution for fixed $\alpha_s$ with no $k_T$ cut-offs} \\
\indent  It is helpful to begin by recalling the solution of the BFKL or
Lipatov
equation in the fixed coupling case in the absence of cut-offs.  The equation
may be written in the form
\begin{displaymath}
-x \frac{\partial f(x,k^2)}{\partial x} \;\; = \;\; \frac{3\alpha_s}{\pi} k^2
\int^{\infty}_0 \frac{dk^{\prime 2}}{k^{\prime 2}} \left[ \frac{f(x, k^{\prime
2}) - f(x,k^2)}{|k^{\prime 2} - k^2|} + \frac{f(x,k^2)}{(4k^{\prime
4}+k^4)^{\frac{1}{2}}} \right]
\end{displaymath}
\begin{equation}
\noindent \hspace*{-5.5cm} \equiv \;\; K \otimes f ,
\end{equation}
where $f(x,k^2)$, the unintegrated gluon distribution, gives the probability
of finding a gluon in the parent hadron with longitudinal momentum fraction
$x$ and transverse momentum squared $k^2$.  To be precise, $f$ is related to
the more familiar integrated gluon distribution, $g(x,Q^2)$, by
\begin{displaymath}
\int^{Q^2}_0 \frac{dk^2}{k^2} f(x,k^2) \;\; = \;\; xg(x,Q^2) .
\end{displaymath}
The BFKL equation sums up the leading powers of log$(1/x)$, that is it
corresponds to the $LL(1/x)$ approximation.  In the genuine $LL(1/x)$
approximation the strong coupling $\alpha_s$ should be taken as a fixed
parameter.  Although the integration over the transverse momentum squared in
(1) does not contain any cut-off parameters, it should be emphasised that it
is free from both infrared and ultraviolet divergences.  However the solution
contains infrared and ultraviolet singularities which will manifest themselves
as non-trivial anomalous dimension(s); we amplify this comment in the
discussion below (19).

Since the kernel $K$ is scale invariant, the equation can be diagonalized by
the Mellin transform
\begin{equation}
f(x,k^2) \;\; = \;\; \frac{1}{2\pi i} \int^{c+i\infty}_{c-i\infty}
(k^2)^{\omega} \tilde{f}(x,\omega)d\omega
\end{equation}
\begin{equation}
\tilde{f}(x,\omega) \;\; = \;\; \int^{\infty}_0 (k^2)^{-\omega -1}
f(x,k^2)dk^2
\end{equation}
where the integration contour in (2) is specified below.  If we substitute (2)
into (1) and change the variable $k^{\prime 2}$ to $k^2u$, then we obtain
\begin{equation}
-x \frac{\partial\tilde{f}(x,\omega)}{\partial x} \;\; = \;\;
\tilde{K}(\omega) \tilde{f}(x,\omega)
\end{equation}
where
\begin{displaymath}
\tilde{K}(\omega) \;\; = \;\; \frac{3\alpha_s}{\pi} \int^{\infty}_0
\frac{du}{u} \left[ \frac{u^{\omega}-1}{|u-1|} +
\frac{1}{(4u^2+1)^{\frac{1}{2}}} \right]
\end{displaymath}
\begin{equation}
\noindent \hspace*{.5cm} = \;\; \frac{3\alpha_s}{\pi} [2\Psi(1) - \Psi(\omega)
- \Psi(1-\omega)] ,
\end{equation}
where $\Psi$ is the logarithmic derivative of the Euler gamma function:
$\Psi(z) \equiv \Gamma^{\prime}(z)/\Gamma(z)$.  If we substitute the solution
of (4) into (2) we obtain
\begin{equation}
f(x,k^2) \;\; = \;\; \frac{1}{2\pi i} \int^{c+i\infty}_{c-i\infty} d\omega
(k^2)^{\omega} \tilde{f}(x_0,\omega) \left( \frac{x}{x_0}
\right)^{-\tilde{K}(\omega)} ,
\end{equation}
where $x_0$ is chosen small enough to ensure the validity of the Lipatov
equation and yet large enough to see the effects of the evolution as $x$
decreases from $x_0$.  The integration contour in formula (6) should be
located between the left-hand singularities of $\tilde{K}(\omega)$
arising from $\Psi(\omega)$, and the right-hand singularities arising from
$\Psi(1-\omega)$ so that (6) reproduces the iterative solution of the original
equation (1).  It is therefore convenient to choose $c = \frac{1}{2}$.  Indeed
$\tilde{K}(\omega)$ is symmetric about $\omega = \frac{1}{2}$ and, as it
varies along the contour ($\omega = \frac{1}{2} + i\nu$ with $-\infty < \nu <
\infty$), has its maximum value at this point.  Thus, from (6), we see that
the region $\omega \sim \frac{1}{2}$ dominates the small $x$ behaviour of
$f(x,k^2)$.  We therefore write $\omega = \frac{1}{2} + i\nu$ and expand
$\tilde{K}$ about $\nu = 0$
\begin{equation}
\tilde{K}(\mbox{$\frac{1}{2}$} + i\nu) \;\; = \;\; \lambda -
\mbox{$\frac{1}{2}$}\lambda^{\prime\prime} \nu^2 + O(\nu^4)
\end{equation}
with
\begin{equation}
\lambda \;\; = \;\; \frac{3\alpha_s}{\pi} \, 4{\rm log}2
\end{equation}
\begin{equation}
\lambda^{\prime\prime} \;\; = \;\; \frac{3\alpha_s}{\pi} \, 28\zeta (3)
\end{equation}
where the Riemann zeta function $\zeta(3) = 1.202$.  If we also expand the
various terms in the integrand of (6) about $\omega = \frac{1}{2}$ (that is
$\nu = 0)$, then we find the behaviour of the gluon for $x \ll x_0$ is given
by the Gaussian integral
\begin{displaymath}
f(x,k^2) \; \approx \; \left( \frac{x}{x_0}\right)^{-\lambda}
\tilde{f}(x_0,\mbox{$\frac{1}{2}$}) \frac{(k^2)^{\frac{1}{2}}}{2\pi}
\int^{\infty}_{-\infty} d\nu \, {\rm exp}\left(
-\mbox{$\frac{1}{2}$}\lambda^{\prime\prime} {\rm log}
\left(\frac{x_0}{x}\right) \nu^2 + i\nu \, {\rm log}
\left(\frac{k^2}{\bar{k}^2} \right)\right)
\end{displaymath}
\begin{equation}
= \;\; \left( \frac{x}{x_0}\right)^{-\lambda}
\frac{\tilde{f}(x_0,\frac{1}{2})(k^2)^{\frac{1}{2}}}
{[2\pi\lambda^{\prime\prime}{\rm log}(x_0/x)]^{\frac{1}{2}}} \, {\rm exp}
\left(\frac{-{\rm log}^2(k^2/\bar{k}^2)}{2\lambda^{\prime\prime}{\rm
log}(x_0/x)} \right)
\end{equation}
where
\begin{equation}
{\rm log}(\bar{k}^2) \;\; = \;\; i \frac{d}{d\nu} ({\rm log}
\tilde{f}(x_0,\mbox{$\frac{1}{2}$} + i\nu))|_{\nu = 0} .
\end{equation}
We have reproduced the solution of the gluon distribution originally obtained
by Lipatov {\it et al.}\ with its characteristic $x^{-\lambda}$ behaviour with
$\lambda$ given by (8), modulated by a $({\rm log}(1/x))^{-\frac{1}{2}}$
factor.

Formula (10) also displays explicitly the diffusion pattern of the
solution of the BFKL equation, that is a Gaussian distribution in
log$(k^2)$ with a width which grows as (log$(1/x))^{\frac{1}{2}}$ as
$x$ decreases.  The position of the maximum of the Gaussian
distribution (given by log$(\bar{k}^2)$ of (11)), as well as the
normalisation of the solution, is controlled by the boundary
conditions, that is by $f(x_0,k^2)$. The rate of diffusion, however, is
independent of the boundary conditions.

The approximate analytic solution (10) only applies for $x \ll x_0$.
If we were to include the exp$(-\frac{1}{2}A\nu^2)$ contribution in
the expansion of $\tilde{f}(x_0,\omega)$ about $\nu = 0$, then the
$\lambda^{\prime\prime}{\rm log}(x_0/x)$ factors in (10) would become
$\lambda^{\prime\prime}{\rm log}(x_0/x) + A$.  We assume, simply for
the purposes of illustration, that this modified form applies for all
$x \leq x_0$.  Fig.\ 1 shows the width of the Gaussian log $k^2$
distribution of $f(x,k^2)/(k^2)^{\frac{1}{2}}$ as $x$ decreases from
$x_0$.  At $x_0$ the \lq\lq width" is given by the boundary conditions
$f(x_0,k^2)/(k^2)^{\frac{1}{2}}$, though in practice this input
distribution will not have a perfect Gaussian form in log $k^2$.  In
Fig.\ 1 we use dashed curves to emphasize the approximate nature of
the treatment for $x \sim x_0$.  It should be noted also that in a
realistic treatment we find that the gluon distribution $f(x,k^2)$
samples $k^2$ uncomfortably close to the infrared (non-perturbative)
region.

\vspace*{.5cm}
\noindent {\bf (b) Physical examples of diffusion of the BFKL solution in
$k^2$} \\
\indent The diffusion in log $k^2$ with decreasing $x$ is a major problem in
the applicability of the BFKL equation since it can lead to an increasingly
large contribution from the infrared and ultraviolet regions of $k^2$ where
the equation is not expected to be valid.  We may illustrate diffusion using
two physical examples from deep-inelastic electron-proton scattering.

Given the unintegrated gluon distribution $f(x,k^2)$ we can, in principle,
calculate the behaviour of the deep-inelastic structure functions
$F_{2,L}(x,Q^2)$ at small $x$ through the so-called $k_T$ factorization
theorem \cite{CAT1,COLELL}.  Then
\begin{equation}
F_i(x,Q^2) \;\; = \;\; \int \frac{dk^{\prime 2}}{k^{\prime 4}} \int^1_x
\frac{dx^{\prime}}{x^{\prime}} f\left( \frac{x}{x^{\prime}},k^{\prime 2}
\right) F^{(0)}_i(x^{\prime},k^{\prime 2},Q^2)
\end{equation}
with $i = 2,L$.  Symbolically we may write $F = f \otimes F^{(0)}$, see Fig.\
2(a), where $f$ describes the gluon ladder and $F^{(0)}$ the quark-box
amplitude for gluon-virtual photon fusion.  It should be noted that the
integration over  $k^{\prime 2}$ extends down to $k^{\prime 2} = 0$ and so
knowledge of $f(x/x^{\prime},k^{\prime 2})$ in this region is, in principle,
necessary for getting absolute predictions for $F_{2,L}(x,Q^2)$.  We return to
this important point in Section 3.

To illustrate the effect of diffusion in $k^2$ we use the $LL(1/x)$
approximation to simplify (12) to
\begin{equation}
F_i(x,Q^2) \;\; = \;\; \int \frac{dk^2}{k^4} f(x,k^2) \, B_i(k^2,Q^2)
\end{equation}
where the \lq\lq impact factors" $B_i$ are
\begin{equation}
B_i(k^2,Q^2) \;\; = \;\; \int^1_0 \frac{dx^{\prime}}{x^{\prime}}
F^{(0)}_i(x^{\prime},k^2,Q^2) .
\end{equation}
Now we may equally well rewrite the convolution (13) by factorizing at an
intermediate link $x_1$ along the gluon chain in Fig.\ 2(a)
\begin{equation}
F_i(x,Q^2) \;\; = \;\; \int \frac{dk^2}{k^4} f(x_1,k^2) \, f_u\left(
\frac{x}{x_1},k^2\right)
\end{equation}
where $f_u$ is a solution of the BFKL equation but with the boundary
condition fixed at the \lq\lq upper" end of the chain by the quark-box impact
factor $B_i(k^2,Q^2)$.  The diffusion pattern is now determined by boundary
conditions at both ends of the gluon ladder \cite{BARTL}.  To be specific, it
is given by
\begin{displaymath}
f(x_1,k^2) f_u\left( \frac{x}{x_1},k^2\right) /k^2 \; \sim \;
\frac{x^{-\lambda}}{\sqrt{{\rm log}(x_0/x_1) {\rm log}(x_1/x)}}
\end{displaymath}

\begin{equation}
{\rm exp} \left( -\frac{{\rm
log}^2(k^2/\bar{k}^2)}{2\lambda^{\prime\prime}{\rm log} (x_0/x_1)} -
\frac{{\rm log}^2(k^2/\bar{k}^2_u)}{2\lambda^{\prime\prime}{\rm log}(x_1/x)}
\right)
\end{equation}
where $\bar{k}^2_u$ is determined by $B_i(k^2,Q^2)$ and so $\bar{k}^2_u \sim
Q^2$.  The variation of the width of the diffusion pattern, as $x_1$ varies
between $x$ and $x_0$, is sketched in Fig.\ 2(a).  Even for large $Q^2$, the
boundary conditions at $x_0$ mean that the infrared region is penetrated
leading to uncertainty in the predictions for $F_i(x,Q^2)$.

This problem is overcome for deep-inelastic $(x,Q^2)$ events
containing an energetic measured jet $(x_j,k^2_j)$, see Fig.\ 2(b)
[16-20].  We then have a $\delta(k^2-k^2_j)$ distribution at the
\lq\lq bottom" of the gluon ladder and $k^2_j$ can be chosen
sufficiently large such that $f(x,k^2)$ does not diffuse appreciably
into the infrared region for physically accessible values of $x/x_j$,
see Fig.\ 2(b).

\vspace*{.5cm}
\noindent {\bf (c) The cut-off dependence of the BFKL solutions} \\
\indent One way to circumvent the diffusion into the non-perturbative region
of $k^2$ is to (artificially) introduce an infrared cut-off $k^2_0$ in
the integration in the BFKL equation (1), and to explore the
sensitivity of the solutions to reasonable variations of the value of
the cut-off.  In this subsection we show that if we introduce an
infrared (or an ultraviolet) cut-off then the structure of the
solution receives only minor changes, although the normalisation is
altered.  Most importantly the leading small $x$ behaviour
$(x/x_0)^{-\lambda}$ remains intact \cite{COLLAND}.

To investigate the sensitivity of the small $x$ behaviour to the choice of
cut-off, it is useful to transform $f(x,k^2)$ to moment space
\begin{equation}
h(n,k^2) \;\; = \;\; \int^1_0 dz \, z^{n-2} f(x,k^2)
\end{equation}
where $z \equiv x/x_0$.  Then the small $x$ behaviour is controlled by the
leading singularity of $h(n,k^2)$ in the complex $n$ plane.  First let us
recover the structure of solution (10) in the absence of a cut-off.  In this
case the equation, after its diagonalisation, (4), by the Mellin transform,
reduces to a simple algebraic equation in moment space.  To be precise, we
transform the integral version of (4) to moment space and we find an algebraic
equation with solution
\begin{equation}
\tilde{h}(n,\omega) \;\; = \;\;
\frac{\tilde{f}_0(\omega)}{n-1-\tilde{K}(\omega)}
\end{equation}
where $\tilde{f}_0(\omega) \equiv \tilde{f}(x_0,\omega)$, and where
$\tilde{h}$ is related to $h$ of (17) just as in (2)
\begin{equation}
h(n,k^2) \;\; = \;\; \frac{1}{2\pi i} \int^{\frac{1}{2} +
i\infty}_{\frac{1}{2} - i\infty} (k^2)^{\omega} \tilde{h}(n,\omega) \, d\omega
{}.
\end{equation}
It follows from (19) that the leading behaviour of $h(n,k^2)$ as $k^2
\rightarrow 0 \, (k^2 \rightarrow \infty)$ is controlled by the nearest
singularity $\omega_+(n) \, (\omega_-(n))$ which lies to the right (left) of
the contour of integration in the $\omega$-plane, that is
\begin{equation}
h(n,k^2) \; \sim \; (k^2)^{\omega_+} \hspace*{2cm} {\rm as} \;\; k^2
\rightarrow 0
\end{equation}
(and as $(k^2)^{\omega_-}$ as $k^2 \rightarrow \infty$).  These $\omega_{\pm}$
singularities come from the zeros of the denominator in (18), and the values
$\omega_{\pm}$ are equal, by definition, to the anomalous dimensions.  Now, as
we shall see in section 3, gauge invariance requires the driving term in the
Lipatov equation to behave as
\begin{displaymath}
f(x_0,k^2) \; \sim \; k^2    \hspace*{2cm} {\rm as} \;\; k^2 \rightarrow 0 ,
\end{displaymath}
corresponding to a pole at $\omega = 1$ in $\tilde{f}_0(\omega)$.  In other
words the presence of the anomalous dimension $\omega_+$ changes the small
$k^2$ behaviour of $h(n,k^2)$ from $k^2$ to $(k^2)^{\omega_+(n)}$ where
$\omega_+(n) < 1$.  It is in this way that the infrared singularities of the
Lipatov equation manifest themselves.

To determine the values of $\omega_{\pm}(n)$ we recall from (7) that in the
neighbourhood of $\omega = \frac{1}{2}$
\begin{displaymath}
\tilde{K}(\omega) \; \approx \; \lambda + \mbox{$\frac{1}{2}$}
\lambda^{\prime\prime} (\omega - \mbox{$\frac{1}{2}$})^2 .
\end{displaymath}
Thus it follows from (18) that $\tilde{h}(n,\omega)$ has two nearby poles at
\begin{equation}
\omega \;\; = \;\; \mbox{$\frac{1}{2}$} \pm
\sqrt{2(n-1-\lambda)/\lambda^{\prime\prime}} \;\; \equiv \;\; \omega_{\pm}(n)
{}.
\end{equation}
These poles move together and pinch the contour in (19) when $n = 1 + \lambda$
and hence lead to a singularity in $h(n,k^2)$ at this point.  Completing the
contour in the left- or right-hand $\omega$-plane, according as $k^2$ is large
or small, gives
\begin{equation}
h(n,k^2) \; = \;
\frac{(k^2)^{\omega_{\pm}}\tilde{f}_0(\omega_{\pm})}{\omega_+-\omega_-} \; =
\; \frac{\sqrt{\frac{1}{2}\lambda^{\prime\prime}}(k^2)^{\omega_{\pm}}
\tilde{f}_0(\omega_{\pm})}{2\sqrt{n-1-\lambda}} .
\end{equation}
If we insert this result in the inverse relation to (17), fold back the
contour in the $n$ plane to circle the square root branch cut and carry out
the $n$ integration, then we again find the behaviour of the gluon
distribution $f(x,k^2)$ given by (10); namely
\begin{equation}
f \;\; \sim \;\; z^{-\lambda}[{\rm log}(1/z)]^{-\frac{1}{2}}
\end{equation}
for small $z \equiv x/x_0$.

To apply this technique to the solution of the Lipatov equation with a $k^2$
cut-off we express $\tilde{h}(n,\omega)$ as the sum of two components
\begin{equation}
\tilde{h}^+(n,\omega) \;\; = \;\; \int^{k^2_0}_0 dk^2 \, (k^2)^{-\omega-1}
h(n,k^2) ,
\end{equation}

\begin{equation}
\tilde{h}^-(n,\omega) \;\; = \;\; \int^{\infty}_{k^2_0} dk^2 \,
(k^2)^{-\omega-1} h(n,k^2) ,
\end{equation}
cf.\ (3).  The component functions $\tilde{h}^+(\tilde{h}^-)$ therefore have
singularities in the right- (left-)hand $\omega$ plane respectively.  In the
case of an infrared cut-off, the algebraic equation which results from taking
the moments of the integral version of (4) is
\begin{equation}
\tilde{h}^+(n,\omega) + \tilde{h}^-(n,\omega) \;\; = \;\;
\frac{\tilde{f}_0(\omega)}{n-1} +
\frac{\tilde{K}(\omega)\tilde{h}^-(n,\omega)}{n-1}
\end{equation}
and so
\begin{equation}
\tilde{h}^-(n,\omega) \;\; = \;\; \frac{\tilde{f}_0(\omega) -
(n-1)\tilde{h}^+(n,\omega)}{n-1-\tilde{K}(\omega)} ,
\end{equation}
as compared to (18).  In the cut-off case we have an additional constraint.
We must adjust $\tilde{h}^+(n,\omega)$ so that $\tilde{h}^-(n,\omega)$ is free
from singularities (i.e.\ poles) in the right half $\omega$ plane.  In
particular the numerator of (27) must contain a factor
\begin{equation}
\omega - \omega_+(n) \;\; = \;\; \omega - {\textstyle \frac{1}{2}} -
\sqrt{2(n-1-\lambda)/\lambda^{\prime\prime}}
\end{equation}
to cancel the pole at $\omega = \omega_+(n)$.  As a consequence the pinch of
the contour of integration in (19) no longer occurs.  We see that
$\tilde{h}(n,\omega)$ (and hence $h(n,k^2)$) contains a $\sqrt{n-1-\lambda}$
singularity, rather than the $1/\sqrt{n-1-\lambda}$ singularity of the non
cut-off case.  Now when we fold the contour in the $n$ plane round the branch
cut and carry out the $n$ integration we find the behaviour
\begin{equation}
f \; \sim \; z^{-\lambda} [{\rm log}(1/z)]^{-\frac{3}{2}}
\end{equation}
for small $z \equiv x/x_0$.  (Clearly we obtain the same behaviour if we have
an ultraviolet, and no infrared, cut-off -- the roles of $\tilde{h}^-$ and
$\tilde{h}^+$ are simply interchanged.)  Finally we note that the infrared
cut-off eliminates infrared singularities and so $h(n,k^2_T) \sim k^2$ as $k^2
\rightarrow 0$ rather than the anomalous behaviour shown in (20).

So far the leading small $x$ behaviour, $x^{-\lambda}$, has remained intact to
the multilations of the Lipatov equation.  However if we introduce both an
infrared and ultraviolet cut-off then the exponent $\lambda$ becomes cut-off
dependent.

\vspace*{.5cm}
\noindent {\bf (d) Fixed and running $\alpha_s$: the eigenvalue spectrum of
the BFKL kernel} \\
\indent The exponent $\lambda$ controlling the small $x$ behaviour,
$x^{-\lambda}$, of the gluon is given by the maximal eigenvalue of the BFKL
kernel, $\tilde{K}(\omega)$; see (6-8).  If we use a fixed value of $\alpha_s$
in the BFKL equation, (1), then the eigenvalue spectrum is continuous, and
remains so in the presence of either an infrared or ultraviolet cut-off.
Moreover the maximum eigenvalue (the branch point of the cut) does not change.

The situation is different if we introduce both an infrared cut-off $k^2_0$
and an ultraviolet cut off $k^2_{{\rm max}}$\cite{COLLAND}.  Then the
eigenvalue spectrum becomes discrete.  The maximum eigenvalue, $\lambda_{{\rm
max}}$, and the separation between the eigenvalues can be shown to depend on
the quantity $t = k^2_{{\rm max}}/k^2_0$.  For large log $t$ the distance
between the eigenvalues becomes proportional to 1/log $t$ and
\begin{equation}
\lambda_{{\rm max}} \;\; = \;\; \lambda_{{\rm max}}(t = \infty) + O(1/{\rm
log}t) ,
\end{equation}
so in the limit $t \rightarrow \infty$ we do indeed recover the continuous
spectrum.

So far we have considered a fixed value of the coupling, $\alpha_s$, in the
BFKL equation, (1).  There are indications that we should allow the
coupling to run, that is $\alpha_s \rightarrow \alpha_s(k^2)$.  Here we make
this physically reasonable replacement, although, as yet, there is no rigorous
proof.  At present the main argument is that in the strongly ordered $k^2$
limit the BFKL equation will produce the Altarelli-Parisi equation in the
double leading logarithm approximation if we take $\alpha_s = \alpha_s(k^2)$.
The introduction of running $\alpha_s$ has the effect of suppressing the
importance of the ultraviolet cut-off and enhancing the dependence on the
infrared behaviour.  Moreover in this case, even with no ultraviolet cut-off,
the eigenvalue spectrum of $\tilde{K}(\omega)$ is discrete, with
$\lambda_{{\rm max}}$ sensitive to the choice of the infrared cut-off $k^2_0$.

\vspace*{1cm}
\noindent {\large{\bf 3.  Numerical solution of the BFKL equation}}

\vspace*{.5cm}
\noindent {\bf (a) Treatment of the infrared region} \\
\indent We shall use a running coupling $\alpha_s(k^2)$ in the BFKL equation
and so we will need to focus attention on how to deal with the infrared
region.  The simplest procedure \cite{AKMS} is to introduce a cut-off $k^2_0$
(as in section 2(b)) so the BFKL equation becomes
\begin{equation}
-x \frac{\partial f}{\partial x} \; = \; \frac{3\alpha_s(k^2)}{\pi} k^2
\int^{\infty}_{k^2_0} \frac{dk^{\prime 2}}{k^{\prime 2}} \left[
\frac{f(x,k^{\prime 2}) - f(x,k^2)}{|k^{\prime 2}-k^2|} +
\frac{f(x,k^2)}{(4k^{\prime 4} + k^4)^{\frac{1}{2}}} \right]
\end{equation}
where (for simplicity) the same cut-off is used in the real emission term and
in the virtual corrections.  To calculate $F_2$ we would impose the same
cut-off on the convolution integral (12) which occurs in the
$k_T$-factorization theorem.

The above cut-off which completely eliminates the infrared region $k^2_T <
k^2_0$ is rather drastic.  Clearly a better procedure which incorporates this
region, at least in an approximate way, is desirable.
The problem is that the BFKL equation is not expected to be valid when the
gluon momenta enter the non-perturbative region of small $k^2$.  One way to
overcome the problem is to introduce non-perturbative (albeit
phenomenological) gluon propagators which are finite at $k^2 = 0$ \cite{HR,FH}
and hence to eliminate the potential infrared singularities of the solution.
Alternatively we could modify the BFKL equation by explicitly subtracting
the point $k^2 = 0$ \cite{COLELL}.  Or, as we shall do below, we could
explicitly introduce a factor $1-G(k^2)$ into the solution with $G(0) = 1.$
The common feature of these modifications of the infrared region is that the
solution of the Lipatov equation (and the driving term) vanish as $k^2$ as
$k^2 \rightarrow 0$.  This requirement that
\begin{equation}
f(x,k^2) \; \sim \; k^2 \hspace*{1cm} {\rm as} \; k^2 \rightarrow 0
\end{equation}
is a consequence of gauge invariance or to be precise of the colour neutrality
of the probed proton \cite{GLR1,LEVORS}.

In the present paper, we attempt to go beyond the simple $k^2$ cutoff
approximation and to model the low $k^2$ region in a more systematic
fashion. We assume that the small-$k^2$ behaviour of the gluon
distribution is driven by a form factor $G(k^2)$ such that

\begin{equation}
f(x,k^2) \;\; \sim \;\; \mbox{Const.} \: [1-G(k^2)]
\end{equation}
for $k^2\rightarrow0$.  We take
\begin{equation}
1 - G(k^2) \; = \; 1 - \frac{k^2_a}{k^2 + k^2_a} \; = \; \frac{k^2}{k^2 +
k^2_a} ,
\end{equation}
where the parameter $k_a^2$ is related to the radius of the gluonic
form factor of the proton. If this is taken to be of the same
magnitude as the radius characterising the hadronic electromagnetic
form factor then we would have $k_a^2\simeq0.5$ GeV$^2$; however
estimates based on the QCD sum rules prefer a larger value,
$k_a^2\simeq 1-2$ GeV$^2$.

We then proceed by splitting the integration region for real gluon
emission (the term involving $f(x,k'^2)$) in (31) up into two parts, namely
\begin{center}
region(A): 0 to $k_0^2$ \\
region (B): $k_0^2$ to $\infty$. \\
\end{center}
In region (B) the BFKL equation as it stands is taken to hold. In
region (A) we assume that $k_0^2$ is sufficiently small that the
behaviour given in (33) is a good approximation. If we parametrise
$f(x,k'^2<k_0^2)$ in this form, then the integral (A) can be
calculated analytically, and in this way we have a physically
motivated approximation for the infrared contribution to the BFKL
equation (31).  As a further modification to the low-$k^2$ region we
`freeze' the argument of $\alpha_s$ by using $\alpha_s(k^2+a^2)$, with
$a^2=1$ GeV$^2$, in both the evolution equation and in the
factorisation formula (12) used to calculate $F_2$ and $F_L$.

The modified BFKL equation can then be used to evolve the gluon
distribution down in $x$ starting from a suitable input distribution
$f(x_0,k^2)$. This boundary condition must be consistent with
Altarelli-Parisi evolution for large $k^2$, that is
\begin{equation}
f^{{\rm AP}}(x_0,k^2) \;\; = \;\; \left. \frac{\partial (x_0g(x_0,Q^2))}
{\partial {\rm log}Q^2}
\right|_{Q^2=k^2}
\end{equation}
where we take  $g(x_0,Q^2)$ from the Altarelli-Parisi evolution of
MRS partons \cite{RGR}. The boundary condition must also have a small
$k^2$ behaviour consistent with our approximations, so we take
\begin{equation}
f(x_0,k^2) \;\; = \;\; (1 - G(k^2)) f^{{\rm AP}}(x_0,k^2+a^2)
\end{equation}
for $k^2 > k^2_0$, whereas for $k^2 < k^2_0$ we \lq\lq freeze" the
evolution of $f^{{\rm AP}}$ at $k^2_0 + a^2$. The parameter $a^2=1$
GeV$^2$ just softens the low-$k^2$ behaviour of $f^{AP}$ which tends
to be unreliable. The important fact is that the input distributions
approach $f^{{\rm AP}}(x_0,k^2)$ for large $k^2$, as they must, and
also embody a suitable behaviour for smaller $k^2$.

In preparation to see the diffusion in $k^2$ develop as we proceed to
small $x$, we plot the boundary conditions in the form
$f(x_0,k^2)/(k^2)^{\frac{1}{2}}$ as suggested by (10).  Sample
distributions are shown in Fig.\ 3 for different choices of $k^2_a$
and $k^2_0$.  We see the input distributions have an approximate
Gaussian form in log $k^2$.  Fig.\ 4 shows the evolution of the
distribution, for the choice $k^2_a = k^2_0 = 1$ GeV$^2$, as we
proceed to smaller $x$ using (31) and (33).  We see both the diffusion
to large $k^2$ and the $x^{-\lambda}$ type growth.  There is no
diffusion into the infrared region since we impose the
phenomenological form (33).  The diffusion to large $k^2$ is more
apparent in Fig.\ 5 which shows the distributions of Fig.\ 4
normalised to a common value.  Even in the limit of very small $x$,
the rate of this diffusion will differ from that given by the analytic
form (10), since here we are using a running coupling $\alpha_s$.

\vspace*{.5cm}
\noindent {\bf (b) Consistency constraint on $k^2_a$ } \\
\indent The numerical solution $f(x,k^2)$ of the Lipatov equation,
modified as above, is found to be much more sensitive to the choice of
$k_a^2$ than to $k_0^2$, see ref. [25]. Therefore here we choose
$k^2_0 = 1$ GeV$^2$, and we concentrate on investigating the
sensitivity of the results to variations of $k_a^2$.  However as may
be anticipated from the discussion in section 2 it is the magnitude of
$f$, and not the shape, which is particularly sensitive to $k^2_a$.
That is, with decreasing $x$, an $x^{-\lambda}$ behaviour sets in with
a numerical value of $\lambda$ only weakly dependent on $k^2_a$
\cite{AKMS2}.

There is a consistency requirement between the \lq\lq input" and \lq\lq
output gluon"
which, in principle, can be used to estimate the value of $k_a^2$.  The
constraint is that the gluon distribution calculated from
\begin{equation}
xg(x,Q^2) \;\; = \;\; \int^{Q^2}_0 \frac{dk^2}{k^2} f(x,k^2) ,
\end{equation}
the inverse relation to (35), should match the phenomenological input
gluon distribution.  The comparison is shown in Fig.\ 6 for different
choices of $k^2_a$.  Here we use as input at $x_0 = 0.01$ a gluon
\cite{RGR} satisfying the leading order Altarelli-Parisi equations
(the dash-dotted curve), but based on the D$_0$ type parametrizations
of ref.\ \cite{MRS1}.  Fig.\ 6 shows that there is good agreement
between the \lq\lq input" and \lq\lq output" gluons for $k^2_a \approx
1$ GeV$^2$. (In Fig.6 we also compare our input gluon with the gluons
of the D$^{\prime}_0$ and D$^{\prime}_-$ next-to-leading order
analyses of ref.\ \cite{MRS2}). In practice, it would be misleading to
impose this constraint too rigorously.  The input gluon is not well
known at $x_0=0.01$, particularly at the low values of $Q^2$ which are
necessary for this comparison.  Nevertheless it is encouraging that
the estimate of $k_a^2$ appears reasonable, and suggests that we
should consider QCD predictions with $k_a^2$ chosen in a range of
about 0.5 to 2 GeV$^2$.

\vspace*{.5cm}
\noindent {\bf (c) Numerical predictions for $F_2$ and $F_L$ at small $x$} \\
\indent We may use the solution $f(x,k^2)$ of the modified Lipatov or BFKL
equation to predict the structure functions $F_i(x,Q^2)$ at small $x$ via the
factorization theorem (12).  The factorization formula has the symbolic form
\begin{equation}
F^{{\rm Lip}}_i \;\; = \;\; f \otimes F^{(0)}_i
\end{equation}
and gives the contribution to $F_i$ arising from the BFKL resummation of soft
gluons.  Recall that $F^{(0)}_i$ describes the quark box (and \lq\lq crossed"
box) photon-gluon fusion process shown in the upper part of Fig.\ 2(a).  The
gluon is off-mass-shell with virtuality approximately equal to $k^{\prime 2}$.
The explicit formula for $F_i^{(0)}(x^{\prime},k^{\prime 2},Q^2)$ can be found
in refs.\ \cite{CAT1,AKMS}.

Before we can obtain a realistic estimate for the structure functions $F_i$ we
must include the background non-Lipatov contributions $F^{{\rm Bg}}_i$.  A
reasonable choice at small $x$ is to assume that $F_i^{{\rm Bg}}$ gradually
increases like $x^{-0.08}$ with decreasing $x$ (as might be expected from a
\lq\lq soft" pomeron with intercept $\alpha_P(0) = 1.08$ \cite{PVL}).  To be
precise we take
\begin{equation}
F^{{\rm Bg}}_i (x,Q^2) \;\; = \;\; F^{{\rm Bg}}_i (x_0,Q^2) (x/x_0)^{-0.08}
\end{equation}
with $x_0 = 0.1$.  The values we use for $F_i^{{\rm Bg}}(x_0,Q^2)$ are listed
in the figure captions.  The resulting predictions for the small $x$
behaviour of $F_i =
F_i^{{\rm Lip}} + F_i^{{\rm Bg}}$ with $i = 2,L$ are shown in Figs.\ 7-10 for
various values of $Q^2$.  In each figure the continuous curves show the
predictions for two choices of the infrared parameter $k^2_a$, namely
$k^2_a=1$ and 2 GeV$^2$.

The recent HERA measurements of $F_2$ are shown on Figs. 7 and 8.  There is
general agreement between QCD and the data.  In particular they both show a
dramatic increase with decreasing $x$, and lie well above a straightforward
extrapolation of the fixed target measurements that exist for $x>10^{-2}$
\cite{NMC,BCDMS}.  Certainly the data indicate  support for the
$x^{-\lambda}$  type behaviour arising from the BFKL leading log$(1/x)$
resummation.  But before we draw conclusions we must consider the effects of
shadowing corrections.

\vspace*{.5cm}
\noindent {\bf (d) Inclusion of shadowing}  \\
\indent The growth of the gluon density with decreasing $x$ means that there
is an increased probability that the gluons will interact and recombine.  To
allow for the effects of this recombination or parton shadowing we incorporate
an additional term in (31)
\begin{equation}
-x \frac{\partial f(x,k^2)}{\partial x} \; = \; K \otimes f -
\frac{81}{16k^2R^2} \alpha^2_s(k^2) [xg(x,k^2)]^2
\end{equation}
where $g$ is given by (37).  The additional term, quadratic in $g$, in (40) is
the leading order shadowing approximation; the negative sign leading to a
suppression in the growth of the gluon density with decreasing $x$.  The
crucial parameter is $R$, where $\pi R^2$ specifies the transverse area in
which the gluons are concentrated within the proton.

For illustration we take $R = 5$ GeV$^{-1}$ (corresponding to gluons uniformly
spread across the proton) and $R = 2$ GeV$^{-1}$ (assuming the gluons are
concentrated in \lq\lq hot-spots" within the proton).   The dashed
curves in Figs.\ 7-10 show the effect of these two shadowing
scenarios respectively on the $k^2_a = 1$ GeV$^2$ prediction.
 We note that the shadowing effects are now slightly stronger than in
 the case \cite{AKMS} when the region
of small $k^{\prime 2}$ was entirely neglected.  However shadowing is still
rather a weak effect in the HERA regime and sets in very gradually, unless
compact \lq\lq hot-spots" of gluons occur.  In particular we are far from the
saturation limit.

We have found \cite{AKMS2} that $F^{{\rm Lip}}_2$ behaves like $Cx^{-\lambda}$
for $x \lapproxeq 10^{-3}$ where the predicted value of $\lambda$ is
relatively insensitive to the uncertainties associated with the infrared
region.  The inclusion of shadowing means that $\lambda$ is no longer constant
but that its value decreases with decreasing $x$, as illustrated by the dashed
curves.  The predictions for $Q^2 =
15$ GeV$^2$ (and $k^2_a = 1$ GeV$^2$) are shown in Fig.\ 11.  Conventional
shadowing ($R$ = 5 GeV$^{-1}$) has relatively little impact on $\lambda$, and
even if \lq\lq hot-spots" were to exist $\lambda$ remains significantly above
the soft-pomeron expectation of 0.08.

We have checked that the large values of $\lambda$, which are essentially
independent of $x$ for $x\leq 10^{-3}$, remain true for other physically
reasonable choices of the background $F_2^{{\rm Bg}}(x,Q^2)$.  Indeed the
values of $\lambda$ obtained in Fig. 11 are much larger than would result from
any straightforward extrapolation of the fixed target $F_2$ data to small $x$.
Consider for example the value of $\lambda$ obtained from the extrapolation of
the $D_0'$ set of MRS partons \cite{MRS2} to small $x$.  (The $D_0$ partons
provide an excellent description of the fixed target data which only exist for
 $x>10^{-2}$).  The parton set $D_0$ develops, via Altarelli-Parisi evolution,
a small $x$ behaviour of the form
\begin{equation}
F_2 \sim \exp[2(\xi(Q^2_0,Q^2){\rm log}(1/x))^{1/2}]
\end{equation}
where the "evolution length"
\begin{equation}
\xi(Q^2_0,Q^2)=\int_{Q_0^2}^{Q^2}{dQ'^{2}\over Q'^{2}}
{3\alpha_s(Q'^{2})\over \pi}
\end{equation}
The increase of $F_2$ with decreasing $x$ can be translated into an effective
$x^{-\lambda}$ behaviour in the HERA regime.  In fact the value of $\lambda$ is
found to be about 0.11 at $Q^2=15$ GeV$^2$ and 0.15 at $Q^2=30$ GeV$^2$;
slightly increasing with $Q^2$ on account of the increase in the evolution
length $\xi(Q^2_0,Q^2)$.  Note that these small values of $\lambda$ rely on the
choice of a sufficiently large value of $Q_0^2$, for instance MRS evolve from
$Q_0^2=4$ GeV$^2$.

     However GRV \cite{GRV} have obtained partons by evolving from a
valence like input at $Q_0^2=0.3$ GeV$^2$.  The very low value of
$Q_0^2$ corresponds to a relatively large evolution length
$\xi(Q^2_0,Q^2)$ for $Q^2$ in the HERA region i.e. $Q^2 \sim 20$
GeV$^2$.  In this way they obtain a steeper small $x$ behaviour for
$F_2$ (see (41) and (42)) compatible with the data.  In fact this
double leading log. behaviour mimics an $x^{-\lambda}$ form with
$\lambda \sim 0.4$ in HERA regime.  However we do not believe that
this is an acceptable explanation of the data since the steepness is
mainly generated in the very low $Q^2$ region where perturbative QCD
is invalid (see, for example, \cite{JRF}).

In order to obtain the steep $x^{-\lambda}$ type behaviour with
$\lambda \sim 0.5$ within the Altarelli-Parisi formalism (and when the
evolution starts at moderately large value of $Q_0^2 \sim 5$ GeV$^2$) one
has to impose this steep behaviour in the parametrisation of the
starting gluon and sea-quark distributions at the reference scale
$Q_0^2$ as it is done for instance in the case of the $D_{-}$ set of
MRS partons \cite {MRS1,MRS2}. In this procedure however one does not
use the BFKL equation, that is the singular behaviour is not generated
explicitly by QCD dynamics.

\vspace*{1cm}
\noindent {\large{\bf 4. Conclusions}}

The recent measurements \cite{H1,ZEUS} of the deep inelastic structure function
$F_2$ at HERA explore the small $x$ regime for the first time.  The data show
that $F_2$ increases as $x$ decreases from $10^{-2}$ to a few
$\times 10^{-4}$, and
do not follow a straightforward extrapolation of the fixed target measurements
\cite {NMC,BCDMS} that exist above $x \sim 10^{-2}$.  This novel behaviour
is in line with the growth anticipated from perturbative QCD via the
$k_T$-factorization formula, symbolically of the form
\begin{equation}
F_2 \;\; = \;\; f \otimes F^{(0)}_2 ,
\end{equation}
which links the small $x$ behaviour of $F_2$ with that of the
universal unintegrated gluon distribution $f$ via the quark box
contribution $F^{(0)}_2$ to photon-gluon fusion. The growth of $F_2$
with decreasing $x$ is thus associated with the BFKL leading
log$(1/x)$ summation of soft gluon emissions which yields the small
$x$ behaviour $f(x,k_T^2)) \sim x^{-\lambda}$ with $\lambda \sim 0.5$.

  However the contribution from the infrared (low $k_T$) region, which
occurs in the convolution of (43), leads to a sizeable uncertainty in
the predictions.  Motivated by the apparent agreement between the data
and the expectations of perturbative QCD, we have attempted to improve
the treatment of the low $k^2_T$ regime.  In particular, rather than
imposing a low $k_T$ cut-off, we employ a physically motivated low
$k_T$ form for $f(x,k^2_T)$ which allows us to extrapolate right down
to $k^2_T = 0$.  This reduces the uncertainty in the predictions for
$F_2$, although the normalisation still depends significantly on the
choice of the value of an infrared (form factor) parameter $k^2_a$.
However the effective slope $\lambda$ which specifies the
$x^{-\lambda}$ shape is much less sensitive to the ambiguities at low
$k_T$.  In Section 2 we gathered together general arguments which
suggested that the slope might be relatively immune from infrared
effects and in Section 3 we performed explicit numerical tests to
verify the result.

In Figs. 7 and 8 we compared the perturbative QCD calculations for $F_2$ with
the recent HERA data.  The dramatic growth with decreasing $x$ is apparent in
both the data and the QCD predictions.  The various curves show the sensitivity
of the QCD determination to the variation of the infrared cut-off and to
shadowing effects.  Fig. 11 translates these results into a comparison for
$\lambda$ [where $\lambda$ is defined by the the $x^{-\lambda}$ growth of the
BFKL component of $F_2(x,Q^2)$].  We see that QCD indeed predicts an
approximate $x^{-0.5}$ behaviour at small $x$ (or in the extreme case of
"hot-spot" shadowing an $x^{-0.3}$ type growth with decreasing $x$).

The results from HERA are very encouraging and suggest that H1 and ZEUS may
have seen the first evidence for the BFKL growth arising from the leading
log$(1/x)$ soft gluon resummation.  Figs. 7,8 and 11 should be regarded as an
indication of what may be learnt when much higher statistics data
become available
and allow a detailed comparison.  At the moment it is only possible
to make predictions
in the leading log$(1/x)$ approximation.  Much effort is being devoted to
obtaining the next-to-leading contributions and, when available, these should
be incorporated.  Inspection of Fig. 11 suggests that, from a study of the $x$
dependence of $\lambda$ for $x\leq 10^{-3}$, we may then be able to quantify
the effects of shadowing.  \\

{\bf Acknowledgments}.

We thank the UK Science and Engineering Research Council for support.  One of
us (JK) thanks the European Community for a "Go-West" Fellowship and another
(ADM) for a "Go-East" Fellowship.  JK is also grateful to Grey College and
Physics Department of the University of Durham for warm hospitality.

\newpage

\noindent {\large {\bf Figure Captions}}

\begin{itemize}
\item[Fig.\ 1] The variation in the width of the Gaussian log $k^2$
distribution of $f(x,k^2)/(k^2)^{\frac{1}{2}}$ as we evolve down in $x$ below
the starting value $x_0$.

\item[Fig.\ 2]  The upper diagrams show a gluon \lq\lq ladder" contribution to
small $x$ (a) for deep-inelastic scattering and (b) for deep-inelastic
scattering together with an energetic jet.  The quark box factor $F^{(0)}$
implicitly includes the contribution of the crossed box.  The lower sketches
show the variation of the width of the log$k^2$ distributions of (16) as a
function of $x_1$.

\item[Fig.\ 3]  The boundary conditions for $f/(k^2)^{\frac{1}{2}}$ at $x =
0.01$ used to solve the modified BFKL equation, for various choices of $k^2_a$
and $k^2_0$.  ($k^2_0 = 1$ GeV$^2$ where it is not specified).

\item[Fig.\ 4]  The evolution of $f/(k^2)^{\frac{1}{2}}$ as we step down in
$x$ using the modified BFKL equation with $k^2_a = k^2_0 = 1$ GeV$^2$.

\item[Fig.\ 5]  As for Fig.\ 4 but with the distributions normalised to a
common value at $k^2 = 1$ GeV$^2$.

\item[Fig.\ 6]  The self consistency of the gluon at $x = 0.1$.  The
dash-dotted
curve is the input gluon \cite{RGR} and the continuous curves show the output
gluon obtained from $f(x,k^2)$ via (37), where $f$ itself is determined from
the input gluon with different choices of $k^2_a$ (but with $k^2_0 = 1$
GeV$^2$).  The dashed curves, which correspond to the D$^{\prime}_0$ and
D$^{\prime}_-$ gluons of ref.\ \cite{MRS2}, are to illustrate the ambiguity in
the input gluon.

\item[Fig.\ 7]  The perturbative QCD predictions for $F_2(x,Q^2)$ at
$Q^2=15$ GeV$^2$ obtained from the $k_T$ factorisation formula, (12).  The
continuous curves correspond to the infrared parameter $k_a^2=1 $ and 2 GeV$^2$
respectively, with shadowing neglected.  The dashed curves show the suppression
caused by conventional ($R=5$ GeV$^{-1}$) and \lq\lq hot-spot"
($R=2$ GeV$^{-1}$) shadowing
for the choice $k_a^2=1$ GeV$^2$.  The data are from the H1 \cite{H1} and ZEUS
\cite{ZEUS} collaborations.  The background contribution is given by (39) with
$F_2^{{\rm Bg}}(x_0)=0.384$.

\item[Fig.\ 8] As for Fig. 7 but for $Q^2=30$ GeV$^2$ with
$F_2^{{\rm Bg}}(x_0=0.1)=0.391$.

\item[Fig.\ 9]  The curves are as for Fig. 7 but for the longitudinal structure
function $F_L(x,Q^2=15 {\rm GeV}^2)$.
The background contribution is given by (39)
with $F_L^{{\rm Bg}}=0.04$ at $x_0=0.1$ and $Q^2=15$ GeV$^2$.

\item[Fig.\ 10]  As for Fig. 9 but at $Q^2=30$ GeV$^2$.

\item[Fig.\ 11] The effective slope $\lambda$, defined by
$F_2=F_2^{{\rm Bg}}+ Cx^{-\lambda}$ where $F_2^{{\rm Bg}}$ is given by
(39), for various
choices of the infrared parameter $k_a^2$.  The lower two (dot-dashed) curves
show the effect of the conventional ($R=5$ GeV$^{-1}$) and \lq\lq hot-spot"
($R=2$ GeV$^{-1}$) shadowing on the $k_a^2=1$ GeV$^2$ predictions.
The \lq\lq data" points
are calculated from the H1 and ZEUS data shown in Figs. 7 and 8.

\end{itemize}


\vspace*{1in}






\newpage
\begin{thebibliography}{99}

\bibitem{KLF}E.A.\ Kuraev, L.N.\ Lipatov and V.\ Fadin, Zh.\ Eksp.\ Teor.\
Fiz.\ {\bf 72} (1977) 373; Sov.\ Phys.\ JETP {\bf 45} (1977) 199
\bibitem{BL}Ya.Ya.\ Balitskij and L.N.\ Lipatov, Yad.\ Fiz.\ {\bf 28} (1978)
1597; Sov.\ J.\ Nucl.\ Phys.\ {\bf 28} (1978) 822.
\bibitem{LIP1}L.N.\ Lipatov, in \lq\lq Perturbative QCD", ed.\ A.H.\ Mueller
(World Scientific, Singapore, 1989) p.\ 411
\bibitem{BS}J.B.\ Bronzan and R.L.\ Sugar, Phys.\ Rev.\ D{\bf 17} (1978) 585
\bibitem{JAR1}T.\ Jaroszewicz, Acta Phys.\ Polon.\ B{\bf 11} (1980) 965
\bibitem{GLR1}L.V.\ Gribov, E.M.\ Levin and M.G.\ Ryskin, Phys.\ Rep.\ {\bf
100} (1983) 1
\bibitem{CIAF}M.\ Ciafaloni, Nucl.\ Phys.\ {\bf 296} (1988) 49
\bibitem{CATFM}S.\ Catani, F.\ Fiorani and G.\ Marchesini, Phys.\ Lett.\ B{\bf
234} (1990) 339; Nucl.\ Phys.\ B{\bf 336} (1990) 18
\bibitem{MARCH1}G.\ Marchesini, Proc.\ of Workshop \lq\lq QCD at 200 TeV",
Erice, Italy, June 1990, eds.\ L.\ Cifarelli and Yu.L.\ Dokshitzer, Plenum
Press, New York 1992, p.\ 183
\bibitem{LEVORS}E.M.\ Levin, Orsay lectures, LPTPE preprint 91/02 (1991)
\bibitem{PVL}P.V.\ Landshoff, Proc.\ XXVII Rencontre de Moriond (Editions
Frontieres, 1992, ed.\ J.\ Tran Thanh Van); \\
A.\ Donnachie and P.V.\ Landshoff, Phys.\ Lett.\ B{\bf 236} (1992) 227;
Manchester preprint M/C-TH 93/11 (1993)
\bibitem{BARTL}J.\ Bartels and H.\ Lotter, Phys.\ Lett.\ B{\bf 309} (1993) 400
\bibitem{COLLAND}J.C.\ Collins and P.V.\ Landshoff, Phys.\ Lett.\ B{\bf 276}
(1992) 196
\bibitem{CAT1}S.\ Catani, M.\ Ciafaloni and F.\ Hautmann, Phys.\ Lett.\ B{\bf
242} (1990) 91; Nucl.\ Phys.\ B{\bf 366} (1991) 135; S.\ Catani, M.\ Ciafaloni
and F.\ Hautmann, Proc.\ of the Workshop \lq\lq Physics at HERA", DESY,
Hamburg, Germany, October 1992, eds.\ W.\ Buchm\"{u}ller and G.\ Ingelman,
Vol.\ 2 (1992) p690
\bibitem{COLELL}J.C.\ Collins and R.K.\ Ellis, Nucl.\ Phys.\ B{\bf 360} (1991)
3
\bibitem{AHM3}A.H.\ Mueller, J.\ Phys.\ G{\bf 17} (1991) 1443
\bibitem{BBDERK}J.\ Bartels, M.\ Besancon, A.\ De Roeck and J.\ Kurzhoefer,
\lq\lq Measurements of Hot Spots at HERA", Proc.\ of the Workshop \lq\lq
Physics at HERA", DESY, Hamburg, Germany, October 1992, eds.\ W.\
Buchm\"{u}ller and G.\ Ingelman
\bibitem{BLDER}J.\ Bartels, M.\ Loewe and A.\ De Roeck, Z.\ Phys.\ C{\bf 54}
(1992) 635; \\
A.\ De Roeck, Nucl.\ Phys.\ B (Proc.\ Suppl.) {\bf 29}A (1992) 61
\bibitem{WKT}W.K.\ Tang, Phys.\ Lett.\ B{\bf 278} (1992) 363
\bibitem{KMS2}J.\ Kwiecinski, A.D.\ Martin and P.J.\ Sutton, Phys.\ Rev.\
D{\bf 46} (1992) 921; J.\ Kwiecinski, A.D.\ Martin and P.J.\ Sutton, Phys.\
Lett.\ B{\bf 287} (1992) 254; A.D.\ Martin, J.\ Kwiecinski and P.J.\ Sutton,
Nucl.\ Phys.\ B (Proc.\ Suppl.) {\bf 29}A (1992) 67
\bibitem{AKMS}A.J.\ Askew, J.\ Kwiecinski, A.D.\ Martin and P.J.\ Sutton,
Phys.\ Rev.\ D{\bf 47} (1993) 3775
\bibitem{HR}R.E.\ Hancock and D.A.\ Ross, Nucl.\ Phys.\ B{\bf 383} (1992) 575
\bibitem{FH}J.R.\ Forshaw and P.N.\ Harriman, Phys.\ Rev.\ D{\bf 46} 1992)
3778
\bibitem{RGR}R.G.\ Roberts, private communication
\bibitem{AKMS2}A.J.\ Askew, J.\ Kwiecinski, A.D.\ Martin and P.J.\ Sutton,
Durham preprint DTP/93/38, June (1993)
\bibitem{MRS1}A.D.\ Martin, R.G.\ Roberts and W.J.\ Stirling, Phys.\ Rev.\
D{\bf 47} (1993) 867
\bibitem{MRS2}A.D.\ Martin, R.G.\ Roberts and W.J.\ Stirling, Phys.\ Lett.\
B{\bf 306} (1993) 145
\bibitem{AKMS}A.J.\ Askew, J.\ Kwiecinski, A.D.\ Martin and P.J.\ Sutton,
Phys.\ Rev.\ D{\bf 47} (1993) 3775
\bibitem{H1}H1 collaboration: DESY preprint, DESY- 93-117 (1993)
\bibitem{ZEUS}ZEUS collaboration: DESY preprint, DESY- 93-110 (1993)
\bibitem{NMC}NMC collaboration:
P. Amaudruz et al., Phys. Lett. B{\bf295} (1992) 159
\bibitem{BCDMS}BCDMS collaboration:
A.C. Benvenuti et al., Phys. Lett. B{\bf 223} (1989) 485
\bibitem{GRV} M. Gl\H{u}ck, E. Reya and A. Vogt, Z.Phys. C{\bf 53} (1992) 127;
Phys. Lett. B{\bf 306} (1993) 391
\bibitem{JRF}J.R. Forshaw, RAL preprint, RAL-92-073
\end{thebibliography}
\end{document}